\documentclass{article} 
\usepackage{iclr2021_conference,times}

\usepackage{amsmath,amsfonts,bm}

\def\eqref#1{equation~\ref{#1}}

\def\1{\bm{1}}

\DeclareMathAlphabet{\mathsfit}{\encodingdefault}{\sfdefault}{m}{sl}
\SetMathAlphabet{\mathsfit}{bold}{\encodingdefault}{\sfdefault}{bx}{n}

\usepackage{hyperref}
\usepackage{url}

\usepackage{verbatim}

\usepackage{multirow}

\usepackage{tikz}
\usetikzlibrary{arrows}

\usepackage{mathabx}

\usepackage{graphicx}
\usepackage{subcaption}

\title{Direct Prediction of Steady-State Flow Fields in Meshed Domain with Graph Networks}

\author{Lukas Harsch, Stefan Riedelbauch
\\
Institute of Fluid Mechanics and Hydraulic Machinery\\
University of Stuttgart\\
Stuttgart, Germany 70569 \\
\texttt{\{lukas.harsch,stefan.riedelbauch\}@ihs.uni-stuttgart.de}
}

\iclrfinalcopy 
\begin{document}

\maketitle

\begin{abstract}
We propose a model to directly predict the steady-state flow field for a given geometry setup. The setup is an Eulerian representation of the fluid flow as a meshed domain. We introduce a graph network architecture to process the mesh-space simulation as a graph. The benefit of our model is a strong understanding of the global physical system, while being able to explore the local structure. This is essential to perform direct prediction and is thus superior to other existing methods.
\end{abstract}

\section{Introduction}
Using deep learning to model physical problems has become an area of great research interest. Learning the physical equations of a systems can be used as model to control robots \citep{planning_trajectories} or to solve numerical problems \citep{Raissi2018DeepHP, SIRIGNANO20181339}. In case of computational fluid dynamics (CFD), learned simulations can reduce the computational cost and thus improve the engineering process of, in our case, hydraulic components.
During the development process of hydraulic machinery, a large amount of designs has to be simulated and evaluated with respect to their fluid dynamical properties. To speed up this process, stationary simulations are often preferred in the early stage of the design process.
For this reason, we propose a method, which enables fast predictions of the fluid dynamics, focusing on predicting the steady-state flow field.
In recent years, deep learning approaches \citep{10.5555/3305890.3306035, Xie2018tempoGANAT} have shown promising results inferring fluid flow simulations. Data-driven approaches are used to learn the underlying physical behavior of the fluid dynamics from CFD simulations. A generalized solution of the fluid dynamics should be learned to infer the flow field for new problem setups.
A major difference of the algorithms is the way the fluid domain is represented. Modeling complex physical systems in general requires a discretized simulation domain. Mesh-based representations are a common choice to solve the underlying partial differential equations (PDEs) of the system, e.g. the Navier-Stokes equations for CFD. 
The mesh can be structured as a Cartesian grid to store the flow field. So image processing algorithms can be applied \citep{10.1145/2939672.2939738, DBLP:journals/corr/abs-1810-08217}, which has its limitations for complex structured geometries. In common numerical solvers the flow field is represented as a mesh of arbitrary polygons, which is widely used in mechanical engineering. With the success of graph neural networks \citep{47094}, it is possible to use deep learning for modeling such physical problems.
Further, the fluid can be treated as particle-based representation, which is common used in smoothed particle hydrodynamics (SPH) and position based fluids (PBF) for simulating free-surface liquids. Particles can be processed as a point cloud. In \citet{schenck2018spnets}, \citet{Ummenhofer2020Lagrangian} and \citet{sanchezgonzalez2020learning}, this representation is used to model the highly dynamic movement of fluids and the interaction with rigid objects.
In our work, we focus on mesh-based representation on an irregular grid. The combination of a differentiable solver with graph convolutions (GCN) \citep{kipf2017semi} is used in \citet{pmlr-v119-de-avila-belbute-peres20a} for high-resolution predictions on the meshed domain around airfoils. Our method comes without an additional solver to be more independent from domain knowledge which is necessary for the solver. \citet{pfaff2021learning} provides {\scshape{MeshGraphNets}}, which is especially designed for predicting the dynamical states of the system over time. In section \ref{sectionResults} we show, that our method is better suited for steady-state predictions.
\section{Model}
\tikzstyle{int}=[draw, fill=gray!20, minimum size=2em, minimum width=1.75cm]
\tikzstyle{int2}=[draw, minimum width=1.75cm]
\tikzstyle{int3}=[draw, minimum width=2.5cm]
\begin{figure}[t]
	\center
	\begin{tikzpicture}
	[cross/.style={path picture={ 
			\draw[black]
			(path picture bounding box.east) -- (path picture bounding box.west) (path picture bounding box.south) -- (path picture bounding box.north);
	}}]

	\node at (0,0) [int2, rotate=90] (a) {$N^v \times F^v$};
	\node [rotate=90] (input) [above of=a,node distance=0.5cm] {Input};
	\node [int] (a2) [right of=a,node distance=1.5cm] {EdgeConv};
	
	\node [int2, rotate=90] (c) [below of=a2,node distance=1.5cm] {$N^v \times 128$};
	\node [int] (c2) [right of=c,node distance=1.5cm] {EdgeConv};
	\node [int2, rotate=90] (d) [below of=c2,node distance=1.5cm] {$N^v \times 128$};
	\node [int] (d2) [right of=d,node distance=1.5cm] {EdgeConv};
	\node [int2, rotate=90] (d3) [below of=d2,node distance=1.5cm] {$N^v \times 128$};
	\node [draw,circle,cross,minimum width=0.5cm] (e) [right of=d3,node distance=1.0cm] {}; 
	\node [int2, rotate=90] (l) [below of=e,node distance=1.0cm] {$N^v \times 384$};

	\path[->] (a) edge node {} (a2);
	\path[->] (a2) edge node {} (c);
	\path[->] (c) edge node {} (c2);
	\path[->] (c2) edge node {} (d);
	\path[->] (d) edge node {} (d2);
	\path[->] (d2) edge node {} (d3);
	\path[->] (d3) edge node {} (e);
	\path[->] (e) edge node {} (l);
	
	\node (g) [above of=c,node distance=1.15cm] {};
	\node (h) [above of=d,node distance=1.0cm] {};
	
	\node (i) [above of=e,node distance=1.15cm] {};
	\node (ir) [right of=i,node distance=0.1cm] {};
	\node (j) [above of=e,node distance=1.15cm] {};
	\node (k) [above of=e,node distance=1.0cm] {};
	\node (kl) [left of=k,node distance=0.1cm] {};
	
	\node (er) [below of=ir,node distance=1.05cm] {};
	\node (el) [below of=kl,node distance=0.9cm] {};
	
	\draw[->] (c.east)--(g.center)--(ir.center)--(er.north);
	\draw[->] (d.east)--(h.center)--(kl.center)--(el.north);
	
	\node (e2) [right of=l,node distance=1.0cm] {};
	\node (e3) [below of=e2,node distance=1.6cm] {};
	\node at (0,-1.6cm) (e4) [] {};
	\node at (0,-3.2cm) (e5) [] {};

	\node [int3, rotate=90] (l2) [below of=e5,node distance=3cm, align=center] {Global Features\\$1024$};
	\node [int2, rotate=90] (n) [below of=l2,node distance=2cm] {$N^v \times 1024$};	
	\node [draw,circle,cross,minimum width=0.5cm] (o) [right of=n,node distance=1.25cm] {};
	\node [int2, rotate=90] (o2) [below of=o,node distance=1.25cm] {$N^v \times 1408$};
	\node [int] (o3) [right of=o2,node distance=1.75cm] {Decoder};
	\node [int2, rotate=90] (p) [below of=o3,node distance=1.75cm] {$N^v \times 3$};
	\node [rotate=90] (q) [below of=p,node distance=0.5cm] {Output};

	\draw [] (l.south) -- (e2.west) -- (e3.west) -- (e4.west) -- (e5.west);
	\path[->] (e5.west) edge node[above,xshift=0cm] {mlp \{1024\}} node[below,xshift=0cm] {max pooling} (l2.north);
	\path[->] (l2) edge node[above] {repeat} (n);
	\path[->] (n) edge node {} (o);
	\path[->] (o) edge node {} (o2);
	\path[->] (o2) edge node {} (o3);
	\path[->] (o3) edge node {} (p);
		
	\node (r) [below of=c,node distance=1.2cm] {};
	\node (s) [below of=d,node distance=1.05cm] {};
	\node (t) [below of=d3,node distance=1.2cm] {};
	
	\node (u) [above of=o,node distance=2.15cm] {};
	\node (u2) [below of=u,node distance=0.15cm] {};
	\node (v) [left of=u2, node distance=0.1cm] {};
	\node (w) [right of=u2,node distance=0.1cm] {};
	
	\node (ol) [below of=v,node distance=1.9cm] {};
	\node (or) [below of=w,node distance=1.9cm] {};
	
	\draw[->] (c.west)--(r.center)--(v.center)--(ol.north);
	\draw[->] (d.west)--(s.center)--(u.center)--(o.north);
	\draw[->] (d3.west)--(t.center)--(w.center)--(or.north);
	
	\end{tikzpicture}
	\caption{The model architecture to predict the flow field, given the node features of shape~$N^v \times F^v$. \textbf{Top row:} {\scshape{EdgeConv}} layers as local feature extractors. \textbf{Top row:} Pooling, to form an 1D global descriptor. Concatenation of the local and global descriptor and transformation by the decoder. $\oplus$:~concatenation.}
    \label{model}
\end{figure}
The task is a steady-state prediction, so the model should learn to directly predict the final steady-state velocity and pressure fields, given only information about structure of the meshed domain and additional domain information, i.e. node positions $\textbf{u}_i$ and additional quantities $\textbf{n}_i$, respectively. We propose a graph neural network model related to Dynamic Graph CNN (DGCNN) \citep{10.1145/3326362}. Figure~\ref{model} shows the visual scheme of our model architecture.
\subsection{Edge Function}
%
%
The meshed domain can be seen as a bidirectional graph $G = (V,E)$, with nodes $V$ connected by mesh edges $E$. 
Each node $i \in V$ contains the mesh-space coordinate $\textbf{u}_i$, as well as quantities $\textbf{n}_i$ describing the domain, e.g. angle of attack, Mach number or the node type to distinguish between the fluid domain and solid objects like the geometry. In addition, each node is carrying the fluid dynamical quantities $\textbf{q}_i$, which we want to model.
The node feature vector $\mathrm{\textbf{v}}_i = [\mathrm{\textbf{u}}_i,\mathrm{\textbf{n}}_i]$ is defined as the concatenation of the nodes position $\mathrm{\textbf{u}}_i$ and the quantities $\mathrm{\textbf{n}}_i$. The total amount of nodes is defined by $N^v$ with a node feature vector size of $F^v$.
%
The edge feature vectors $\mathrm{\textbf{e}}_{ij}\in E$ are defined as $\mathrm{\textbf{e}}_{ij}=h_\Theta (\mathrm{\textbf{v}}_i,\mathrm{\textbf{v}}_j)$, where $h_\Theta$ is a nonlinear function with a set of learnable parameters~$\Theta$.
Extraction both local and global features is essential for the task of directly predicting a global field, since it is necessary to understand the local properties of the mesh as well as its global structure. Therefore, we are using an edge function called {\scshape{EdgeConv}} introduced by DGCNN, which is defined as $h_\Theta (\mathrm{\textbf{v}}_i,\mathrm{\textbf{v}}_j)=\bar{h}_\Theta (\mathrm{\textbf{v}}_i,\mathrm{\textbf{v}}_j-\mathrm{\textbf{v}}_i)$. It is implemented as
\begin{equation}
    \mathrm{\textbf{e}}'_{ij} = \mathrm{ReLU} (\Theta \cdot (\mathrm{\textbf{v}}_j-\mathrm{\textbf{v}}_i) + \phi \cdot \mathrm{\textbf{v}}_i)\;, \qquad \mathrm{\textbf{v}}'_i = \underset{j:(i,j)\in E}{\mathrm{max}} \mathrm{\textbf{e}}'_{ij}\;,
\end{equation}
with $\Theta$ and $\phi$ implemented as a shared MLP and $\mathrm{max}$ as aggregation operation.

The local neighborhood information are captured by $(\mathrm{\textbf{v}}_j-\mathrm{\textbf{v}}_i)$, which includes the relative displacement vector in mesh space $\mathrm{\textbf{u}}_{ij}=\mathrm{\textbf{u}}_i-\mathrm{\textbf{u}}_j$ and the difference of the local quantities $\mathrm{\textbf{n}}_i$. The global shape structures are captured by the patch center $\mathrm{\textbf{v}}_i$ including the absolute node position $\mathrm{\textbf{u}}_i$ and its quantities $\mathrm{\textbf{n}}_i$. The MLP is embedding the feature vector $\mathrm{\textbf{v}}_i$ to a latent vector of size 128. The {\scshape{EdgeConv}} formulation has properties lying between translation-invariance and non-locality. More details about this model type can be found in \citet{10.1145/3326362}.
\subsection{Architecture}
The DGCNN architecture was originally designed for point clouds, using a k-nearest neighbor algorithm to construct the graph. Since the simulation domain is already present as a mesh we rely on the graph structure of the mesh instead of recomputing the graph.
We use the structure of the DGCNN architecture to increase the ability of extracting local and global properties as displayed in Figure~\ref{model}. The model has $L$ layers of {\scshape{EdgeConv}}. The outputs of the {\scshape{EdgeConv}} layers can be seen as local feature descriptors. Shortcut connections are included to extract multi-scale features. The features from all {\scshape{EdgeConv}} layers are concatenated and aggregated by a fully-connected layer. Then, a global max pooling is used to get the global feature vector of the mesh. Afterwards, the global feature vector is repeated and concatenated with the local features.  At last, the decoder MLP is transforming the latent features to the desired output $\mathrm{\textbf{p}}_i$. Further information about the network and hyperparameter settings can be found in Section \ref{modelDetails}.
%
%
%
%
\begin{figure}[t]
    \begin{subfigure}{.49\textwidth}
        \centering
        \includegraphics[width=0.49\linewidth]{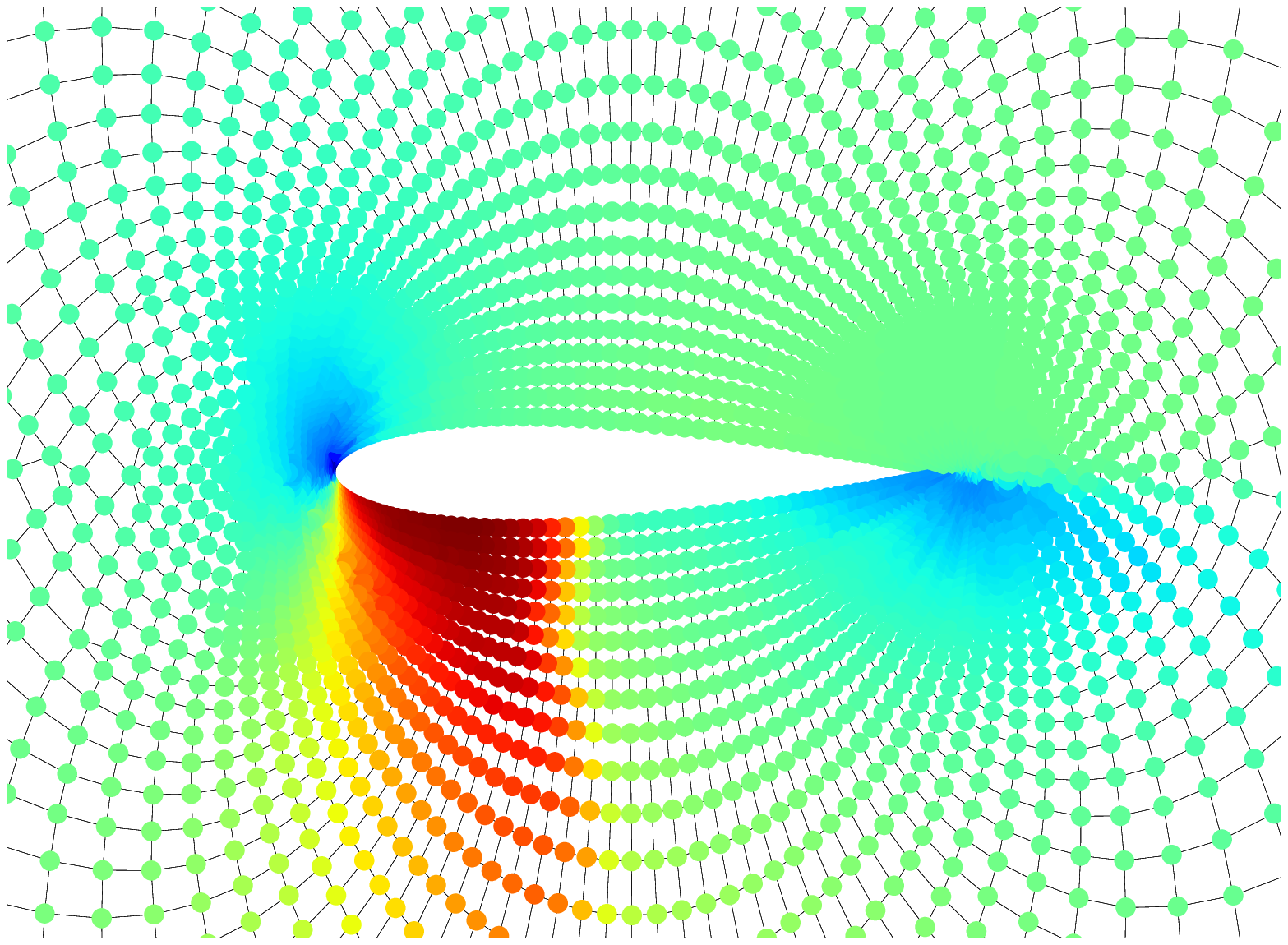}
        \includegraphics[width=0.49\linewidth]{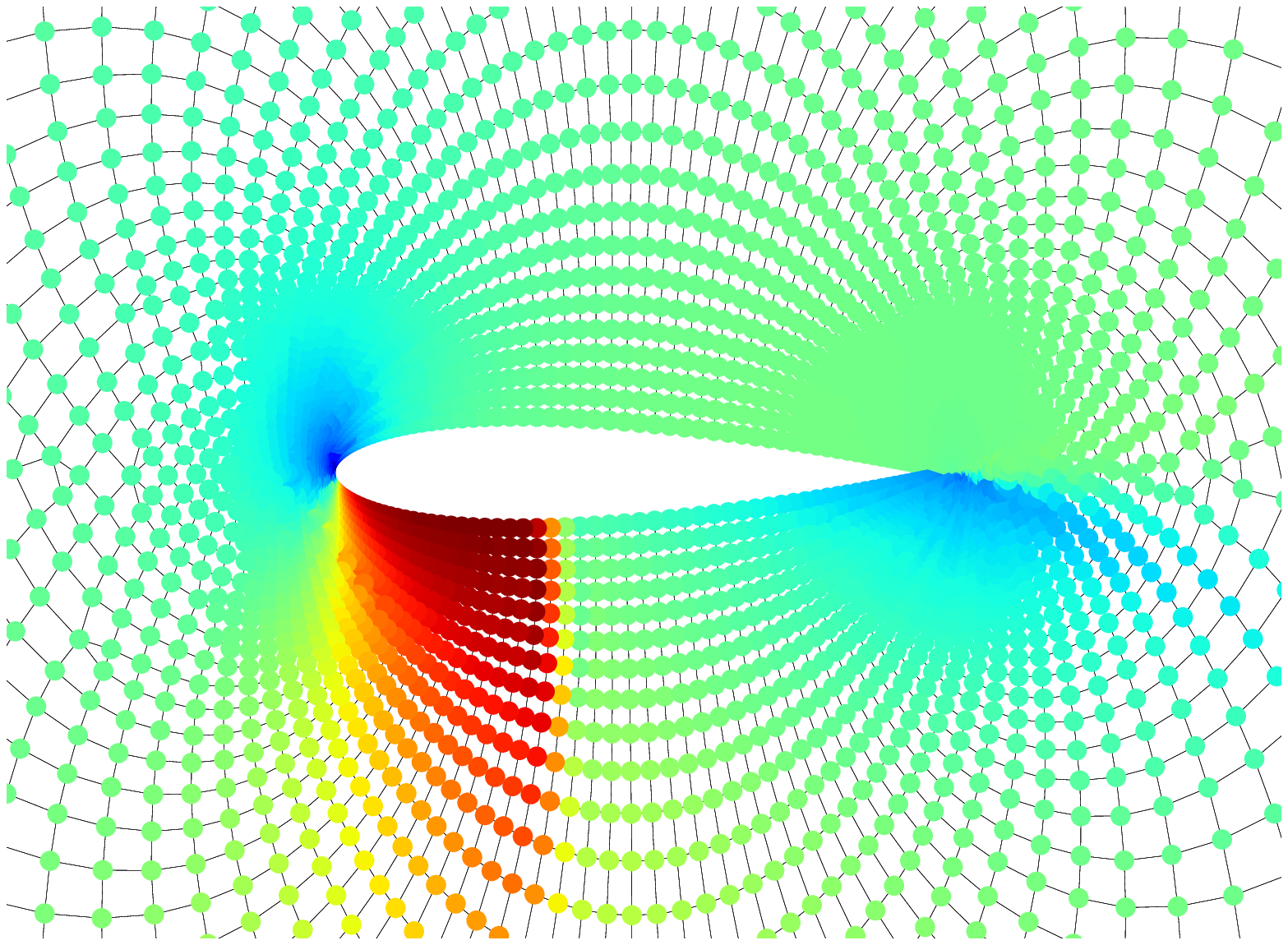}
        \caption{{\scshape{Airfoil}}}
    \end{subfigure}%
    \hfill
    \begin{subfigure}{.49\textwidth}
        \centering
        \includegraphics[width=0.49\linewidth]{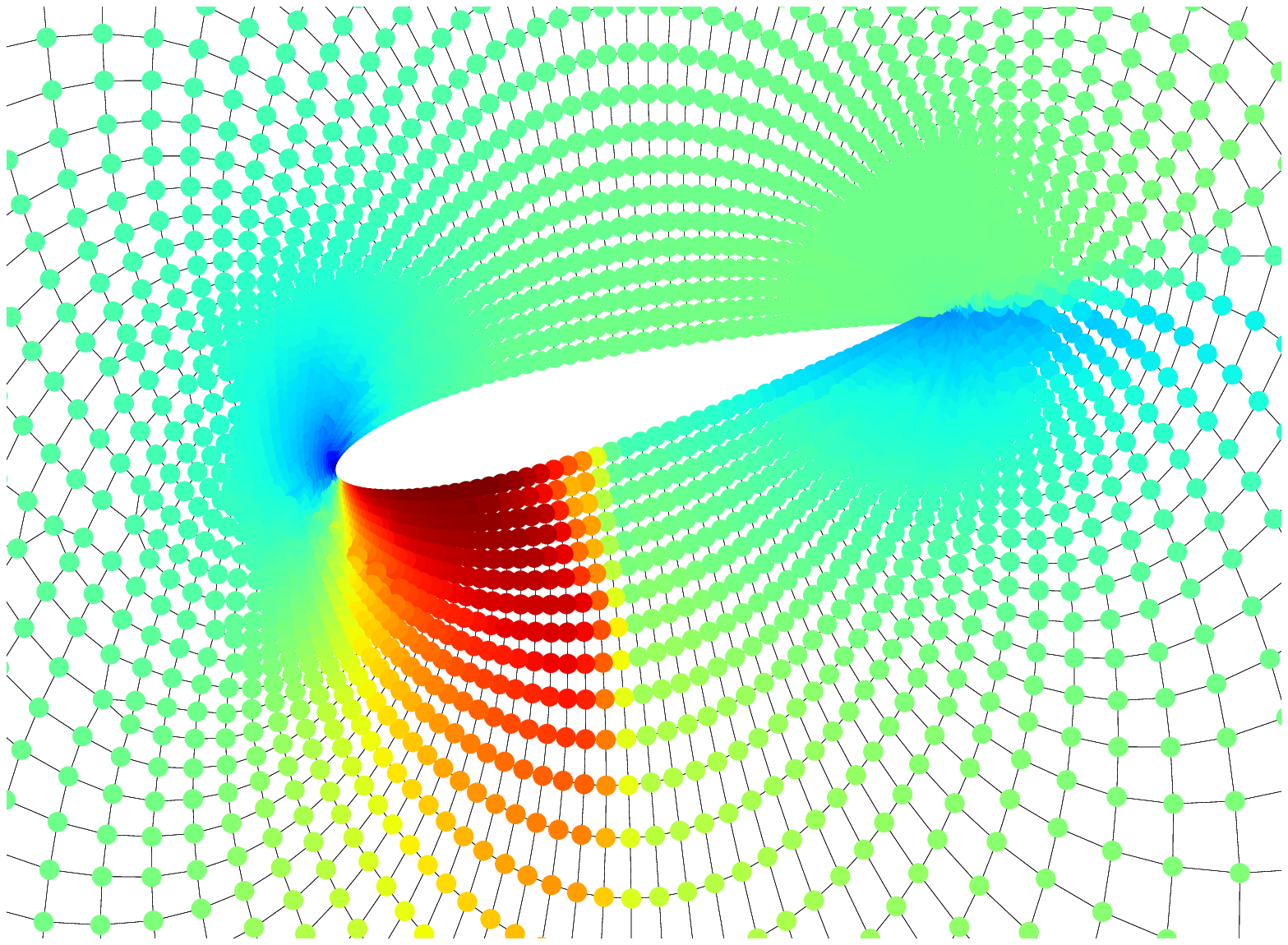}
        \includegraphics[width=0.49\linewidth]{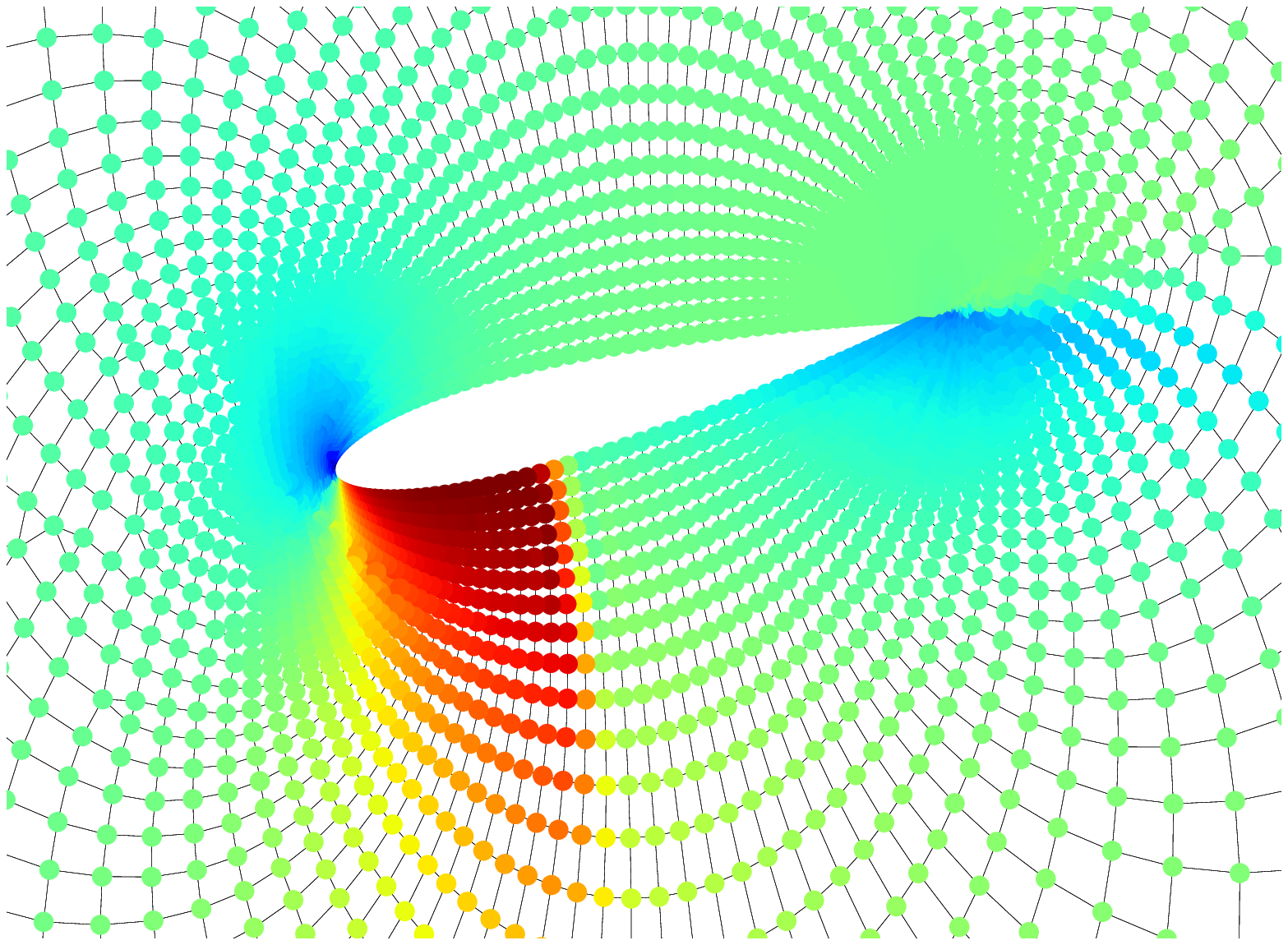}
        \caption{{\scshape{AirfoilRot}}}
    \end{subfigure}%
    \\
    \begin{subfigure}{1\textwidth}
        \centering
        \includegraphics[width=0.49\linewidth]{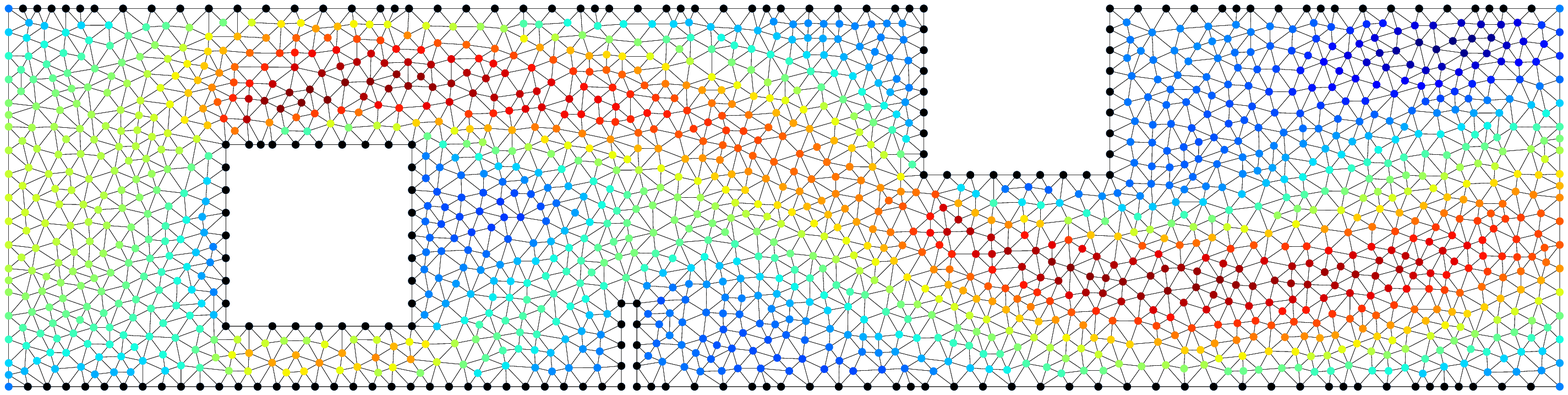}
        %
        %
        \includegraphics[width=0.49\linewidth]{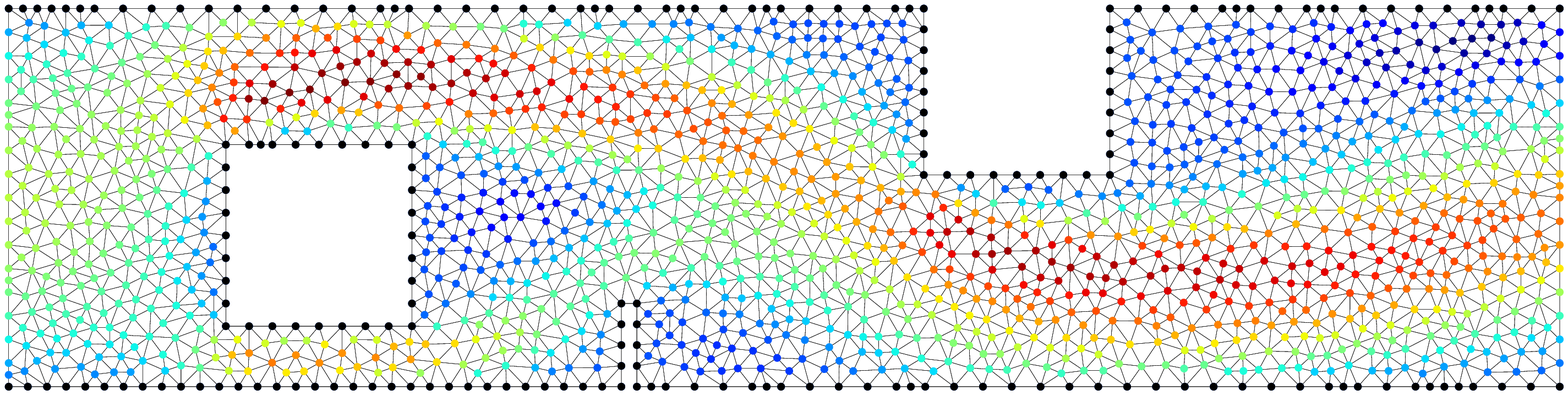}
        \caption{{\scshape{Channel}}}
    \end{subfigure}
    \caption{Examples of the predicted velocity field in $x$-direction for all three data sets, each with the ground truth simulation to the right. The simulation mesh is displayed in the background. For {\scshape{Channel}}, the geometry and walls are highlighted in black.}
    \label{resultsFlowField}
\end{figure}
\section{Experimental setup}
%
%
%
%
%
For our experiments, three data sets with fluid dynamical simulations on two-dimensional meshes are used.
The flow field is represented by the simulation mesh. Each node is holding the fluid dynamical quantities $\mathrm{\textbf{q}}_i=[\mathrm{\textbf{w}}_i,p_i]$, with the velocity in $x$- and $y$-direction $\mathrm{\textbf{w}}_i$ and the pressure~$p_i$.
All samples in the data set {\scshape{Airfoil}} are steady-state, compressible, inviscid CFD simulations, performed by solving the Euler equations. 
%
The same simulation conditions apply to data set {\scshape{AirfoilRot}}, which is a modified version of {\scshape{Airfoil}}.
The CFD simulations of the data set {\scshape{Channel}} are performed by solving the steady-state, incompressible, inviscid case of the Navier-Stokes equations.
The first data set {\scshape{Airfoil}} is taken from \citet{belbute_peres_cfdgcn_2020}. The simulations are describing the flow around the cross-section of a NACA0012 airfoil for varying angle of attacks $\alpha$ and different Mach numbers $m$. It includes a single geometry setup, so the mesh is identical for every sample in the data set. Thus, the simulated quantities $\mathrm{\textbf{q}}_i$ vary only due to the parameter setting of $\alpha$ and $m$. 
The node feature vectors $\mathrm{\textbf{v}}_i$ include the node positions $\mathrm{\textbf{u}}_i$, the global $m$ and $\alpha$ and the node type, indicating, whether the node belongs to the airfoil geometry, the fluid domain or the boundary of the flow field. 
Since the node positions are the same for every simulation, the model predictions mainly depend on the parameter setup of $\alpha$ and $m$. Consequently, it is difficult to show the advantage of our method to capture local and global structures. 
This data set is used as a baseline to compare our method with the results from literature.
A second data {\scshape{AirfoilRot}} set is constructed, by rotating the whole simulation domain of {\scshape{Airfoil}} by the value of $\alpha$. This can be interpreted as $\alpha=0$, but with a rotated placement of the simulation mesh and its flow field. Here, $\alpha$ will be discarded from the feature vector, since it is equal for all samples. Now, the model has to rely more on the geometric structure for predicting the flow field.
All samples in the data set still have the same geometry profile. Nevertheless, the orientation of the airfoil results in strong variations of the flow field, so the model has to be sensitive in understanding the geometric structure.
%

%
To increase the the requirement of understanding the geometric structure, we generate a third data set {\scshape{Channel}}, which is a two-dimensional channel flow. Generic objects like circles, squares, triangles and rectangle are randomly placed inside the channel. The number of objects varies between one and three, which results in a large variety of fluid dynamics i.e., various flows, back flows, partial blockage and small gaps. The feature vector for {\scshape{Channel}} only consists of the node positions $\mathrm{\textbf{u}}_i$ and the node type, i.e. fluid domain, wall or internal object. All samples in {\scshape{Channel}} are simulated with the same boundary conditions, so there is no additional information encoded in the feature vector. Thus, the focus lies on the investigation for the understanding of the geometric structure.
Examples for all three data sets are shown in Figure \ref{resultsFlowField}. More details about the data sets are explained in Section \ref{dataDetails}.

\section{Results} \label{sectionResults}
\begin{table*}[t]
	\center
	\caption{Test root mean square error (RMSE) for all tasks (RMSE$\times 10^{-2}$).}
	\label{resultsTable}
	\begin{tabular}{|l||c|c|c|}
		\hline
		\rule{0pt}{0.35cm} 
		Method&{\scshape{Airfoil}}&{\scshape{AirfoilRot}}&{\scshape{Channel}}\\
		\hline
		\hline
		\rule{0pt}{0.35cm} 
		GCN&1.40&-&-\\
		\hline
		\rule{0pt}{0.35cm} 
		{\scshape{MeshGraphNets}}&0.95&7.34&13.30\\
		\hline
		\rule{0pt}{0.35cm} 
		{\scshape{MeshGraphNets} (Abs)}&\textbf{0.73}&1.28&13.04\\
		\hline
		\rule{0pt}{0.35cm} 
		{\scshape{MeshGraphNets} (Abs+Pool)}&1.25&0.99&6.58\\
		\hline
		\rule{0pt}{0.35cm} 
		Ours&1.40&\textbf{0.87}&\textbf{5.81}\\
		\hline
	\end{tabular}
\end{table*}
We compare our approach with two baseline methods, a base GCN and {\scshape{MeshGraphNets}} \citep{pfaff2021learning}. The results for the base GCN are taken from \citet{pmlr-v119-de-avila-belbute-peres20a}. They are only available for the {\scshape{Airfoil}} data set. Then, we use a re-implementation of {\scshape{MeshGraphNets}} for comparison with the data sets studied in this work.
{\scshape{MeshGraphNets}} was designd for modeling the system dynamics over time. The models spatial equivariance bias is counteracting the direct prediction of a final flow field.
However, the model can achieve good results for {\scshape{Airfoil}}, but fails to predict the flow field for the other data sets.
We investigate two extensions of {\scshape{MeshGraphNets}}, indicated by the suffix {\scshape{(Abs)}} and {\scshape{(Abs+Pool)}}. First, the absolute positions $\mathrm{\textbf{u}}_i$ is concatenated to the node features {\scshape{(Abs)}}. Second, we include the pooling operation for global feature extraction {\scshape{(Abs+Pool)}}. Both extensions are adopted from the DGCNN architecture and aim to increase the understanding of the global structure. The adapted model is described more detailed in Section \ref{adaptionDetails}.
The visual results of our method are displayed in Figure \ref{resultsFlowField}, showing the predicted and the ground truth field of the velocity component in $x$-direction. There is almost no visually noticeable difference between the predicted and the ground truth flow field. This shows, that our model is able to produce high-quality predictions. More visual results of the velocity fields are shown in Figure \ref{resultsDetails}.
%

%
Table \ref{resultsTable} shows the prediction error in terms of RMSE for all five models on the three data sets. 
In the simplest data set {\scshape{Airfoil}}, with respect to the complexity of the geometric structure, all models successfully learn to predict the flow field. As mentioned in \citet{pfaff2021learning} {\scshape{MeshGraphNets}} outperform the base GCN, which is also the case with our re-implementation. Surprisingly, {\scshape{(Abs)}} outperforms all the other methods. In terms of RMSE, our method falls behind, but visually the fluid dynamics remain nearly identical to the ground truth.
On the richer {\scshape{AirfoilRot}} task, {\scshape{MeshGraphNets}} was unable to obtain a correct prediction of the flow field.
This results in a high RMSE, which is also confirmed by visual investigations. In contrast, the adapted {\scshape{MeshGraphNets}} capture enough geometric information to successfully predict the flow field. This experiment shows the benefit of using both the local displacement and the absolute node position as features. In this case, our model outperforms the other methods in terms of RMSE. 
Beside the velocities and pressure, the flow field of a compressible flow should also be represented by the density, but {\scshape{Airfoil}} and {\scshape{AirfoilRot}} do not provide values for the density. Nevertheless, including the density as an additional output quantity will increase the complexity of the problem setup and thus, increase the prediction error slightly. 
Further, it has to be mentioned, that the test set of {\scshape{Airfoil}} and {\scshape{AirfoilRot}} is within the Mach number range of the training set. So the test case is an interpolation task. Since we are not using any additional physical equations or numerical solvers, our model fails to solve difficult extrapolation tasks as presented in \citet{pmlr-v119-de-avila-belbute-peres20a}. 
For the most challenging data set {\scshape{Channel}}, also {\scshape{(Abs)}} fails. Only {\scshape{(Abs+Pool)}} and our method achieve a sufficient low RMSE, but with an improvement of the RMSE with our method. The same findings are indicated by visual investigations of the predicted flow fields. This experiment shows, that our method generalizes well to new unseen geometric setups of the channel flow. In this experiment, both models with a global feature extractor are successful. This shows, that extracting the global features with the pooling operation is necessary to capture a high level of understanding of the global geometric structure. Nevertheless, our method is better in terms of model complexity and processing time, which makes it superior to {\scshape{MeshGraphNets (Abs+Pool)}}.
\section{conclusion}
We propose a mesh-based method to model the fluid dynamics for an accurate and efficient prediction of steady-state flow fields. The experiments demonstrate the models strong understanding of the geometric structure. The method is independent of the structure of the simulation domain, being able to capture highly irregular meshes. 
Further, our method works well for different kind of systems, i.e. Eulerian and Navier-Stokes systems.
Beside this, we show, that the model does not require any a-priori domain information, e.g. inflow velocity or material parameters. Thus, the model can be used for other systems, e.g.  viscous fluids or other domains which are represented as field data like rigid body simulations.
%
%
%
\bibliography{iclr2021_conference}
\bibliographystyle{iclr2021_conference}
\appendix
\section{Appendix}
\subsection{Additional Model Details} \label{modelDetails}
All MLPs in the {\scshape{EdgeConv}} layers are using three shared fully-connected layers of size 128. The two hidden layers come with ReLU activation, followed by a non-activated output layer. The MLPs in the {\scshape{EdgeConv}} layers are followed by a BatchNorm \citep{pmlr-v37-ioffe15} layer. To aggregate the features of all {\scshape{EdgeConv}} layers, a fully-connected layer of size 1024 is used. The decoder MLP has four fully-connected layers of size (1024, 512, 256, 3) to transform the latent feature vector to the output shape. We are using $L=5$ {\scshape{EdgeConv}} layers, which is a good trade-off between model capacity and computational cost. The whole model consists of 20 fully-connected layers with a total number of approximately 3.3~million parameters.

We train our model in a supervised fashion using the per-node output features $\mathrm{\textbf{p}}_i$ from the decoder with a $L_2$ loss between $\mathrm{\textbf{p}}_i$ and the corresponding ground truth values $\mathrm{\textbf{p}}_i$. A batch size of 2 is chosen for {\scshape{Airfoil}} and {\scshape{AirfoilRot}}. In {\scshape{Channel}} the number of nodes is smaller compared to the former data sets, so a batch size of 6 is used. The models are trained on a single RTX 3090 GPU with the Adam optimizer, using a learning rate of $10^{-4}$.
\subsection{Dataset} \label{dataDetails}
%
The simulation domain of {\scshape{Airfoil}} and {\scshape{AirfoilRot}} is a two-dimensional, mixed triangular and quadrilateral mesh with 6648 nodes. The mesh is highly irregular with varying scale of the edge lengths in different regions of the mesh. The training set is defined by
\begin{equation*}
    \alpha_{\mathrm{train}}=\{-10,-9,\ldots,9,10\}\;, \qquad m_{\mathrm{train}}=\{0.2,0.3,0.35,0.4,0.5,0.55,0.6,0.7\}\;.
\end{equation*}
The test set is defined similar by
\begin{equation*}
    \alpha_{\mathrm{test}}=\{-10,-9,\ldots,9,10\}\;, \qquad m_{\mathrm{test}}=\{0.25,0.45,0.65\}\;.
\end{equation*}
The training and test pairs are sampled uniformly from $\alpha \times m$. This results in 168 samples for the training set and 63 samples for the test set, which is split to 32 samples for validation and 31 samples for the test. The samples from {\scshape{Airfoil}} and {\scshape{AirfoilRot}} come with a min-max normalization to the value range of $[-1,1]$ separately for each velocity component and the pressure. 
%

%
The simulation domain of {\scshape{Channel}} is a two-dimensional, unstructured triangular mesh with an equal scale of the edge lengths. Depending on shape and number of objects inside the channel, the number of nodes is variable, but is approximately 1750. This data set has 1000 samples for training and 150 samples each for validation and test. In {\scshape{Channel}}, we normalize the samples by the maximum velocity and the maximum pressure.
The node features used as inputs as well as the outputs are summarized in the following table.
\begin{table*}[h]
	\center
	\begin{tabular}{|l||c|c|}
		\hline
		\rule{0pt}{0.35cm} 
		Method&inputs $\mathrm{\textbf{v}}_i$&outputs $\mathrm{\textbf{p}}_i$\\
		\hline
		\hline
		\rule{0pt}{0.35cm} 
		{\scshape{Airfoil}}&$\mathrm{\textbf{u}}_i$, $\mathrm{n}_i$, $m$, $\alpha$&$\mathrm{\textbf{w}}_i,p_i$\\
		\hline
		\rule{0pt}{0.35cm} 
		{\scshape{AirfoilRot}}&$\mathrm{\textbf{u}}_i$, $\mathrm{n}_i$, $m$&$\mathrm{\textbf{w}}_i,p_i$\\
		\hline
		\rule{0pt}{0.35cm} 
		{\scshape{Channel}}&$\mathrm{\textbf{u}}_i$, $\mathrm{n}_i$&$\mathrm{\textbf{w}}_i,p_i$\\
		\hline
	\end{tabular}
\end{table*}
The inputs are defined by the node position $\mathrm{\textbf{u}}_i$, the node type $\mathrm{n}_i$, the angle of attacks $\alpha$ and the Mach numbers $m$, while the outputs are the velocity in $x$- and $y$-direction $\mathrm{\textbf{w}}_i$ and the pressure $p_i$.
%
%
%
%
%
%
\subsection{MeshGraphNets Adaption} \label{adaptionDetails}
\tikzstyle{int}=[draw, fill=gray!20, minimum size=2em, minimum width=3cm]
\tikzstyle{int2}=[draw, minimum width=3.5cm]
\tikzstyle{int3}=[draw, minimum width=2.5cm]
\begin{figure}[t]
	\center
	\begin{tikzpicture}
	[cross/.style={path picture={ 
			\draw[black]
			(path picture bounding box.east) -- (path picture bounding box.west) (path picture bounding box.south) -- (path picture bounding box.north);
	}}]
	
	\node at (0,0) [int2, rotate=90] (a) {$N^v \times F^v$, $N^e \times F^e$};
	\node [rotate=90] (input) [above of=a,node distance=0.5cm] {Input};
	\node [int, rotate=90] (a2) [below of=a,node distance=1.25cm] {Encoder};
	\node [int2, rotate=90] (a3) [below of=a2,node distance=1.25cm] {$N^v \times 128$, $N^e \times 128$};
	\node [int, rotate=90] (a4) [below of=a3,node distance=1.25cm] {Message Passing};
	
	\node [int2, rotate=90] (b) [below of=a4,node distance=1.25cm] {$N^v \times 128$, $N^e \times 128$};
	\node [int, rotate=90] (b2) [below of=b,node distance=1.25cm] {Message Passing};
	\node [int2, rotate=90] (c) [below of=b2,node distance=1.25cm] {$N^v \times 128$, $N^e \times 128$};
	\node [int, rotate=90] (c2) [below of=c,node distance=1.25cm] {Message Passing};
	\node [int2, rotate=90] (d) [below of=c2,node distance=1.25cm] {$N^v \times 128$, $N^e \times 128$};
	\node [draw,circle,cross,minimum width=0.5cm] (e) [right of=d,node distance=1.25cm] {}; 
	
	\path[->] (a) edge node {} (a2);
	\path[->] (a2) edge node {} (a3);
	\path[->] (a3) edge node {} (a4);
	\path[->] (a4) edge node {} (b);
	\path[->] (b) edge node {} (b2);
	\path[->] (b2) edge node {} (c);
	\path[->] (c) edge node {} (c2);
	\path[->] (c2) edge node {} (d);
	\path[->] (d) edge node {} (e);
	
	\node (f) [above of=b,node distance=2.15cm] {};
	\node (g) [above of=c,node distance=2cm] {};
	\node (h) [above of=d,node distance=1.85cm] {};
	
	\node (i) [above of=e,node distance=2.15cm] {};
	\node (ir) [right of=i,node distance=0.1cm] {};
	\node (j) [above of=e,node distance=2cm] {};
	\node (k) [above of=e,node distance=1.85cm] {};
	\node (kl) [left of=k,node distance=0.1cm] {};
	
	\node (er) [below of=ir,node distance=2.05cm] {};
	\node (el) [below of=kl,node distance=1.75cm] {};
	
	\draw[->] (b.east)--(f.center)--(ir.center)--(er.north);
	\draw[->] (c.east)--(g.center)--(j.center)--(e.north);
	\draw[->] (d.east)--(h.center)--(kl.center)--(el.north);
	
	\node (e2) [right of=e,node distance=0.75cm] {};
	\node (e3) [below of=e2,node distance=2.6cm] {};
	\node at (0,-2.6cm) (e4) [] {};
	\node at (0,-4.6cm) (e5) [] {};
	
	\node [int2, rotate=90] (l) [below of=e5,node distance=0.55cm] {$N^v \times 640$};
	\node [int3, rotate=90] (l2) [below of=l,node distance=3cm, align=center] {Global Feature\\$1024$};
	\node [int3, rotate=90] (n) [below of=l2,node distance=2cm] {$N^v \times 1024$};	
	\node [draw,circle,cross,minimum width=0.5cm] (o) [right of=n,node distance=1.5cm] {};
	\node [int3, rotate=90] (o2) [below of=o,node distance=1.5cm] {$N^v \times 1664$};
	\node [int, rotate=90] (o3) [below of=o2,node distance=1.5cm] {Decoder};
	\node [int3, rotate=90] (p) [below of=o3,node distance=1.5cm] {$N^v \times 3$};
	\node [rotate=90] (q) [below of=p,node distance=0.5cm] {Output};
	
	\draw [->] (e.west) -- (e2.west) -- (e3.west) -- (e4.west) -- (e5.west) --(l.north);
	\path[] (l.south) edge node[above,xshift=0cm] {mlp \{1024\}} node[below,xshift=0cm] {max pooling} (l2.north);
	\path[->] (l2) edge node[above] {repeat} (n);
	\path[->] (n) edge node {} (o);
	\path[->] (o) edge node {} (o2);
	\path[->] (o2) edge node {} (o3);
	\path[->] (o3) edge node {} (p);
	
	\node (r) [below of=b,node distance=2cm] {};
	\node (s) [below of=c,node distance=1.85cm] {};
	\node (t) [below of=d,node distance=2cm] {};
	
	\node (u) [above of=o,node distance=2.75cm] {};
	\node (u2) [below of=u,node distance=0.15cm] {};
	\node (v) [left of=u2, node distance=0.1cm] {};
	\node (w) [right of=u2,node distance=0.1cm] {};
	
	\node (ol) [below of=v,node distance=2.5cm] {};
	\node (or) [below of=w,node distance=2.5cm] {};
	
	\draw[->] (b.west)--(r.center)--(v.center)--(ol.north);
	\draw[->] (c.west)--(s.center)--(u.center)--(o.north);
	\draw[->] (d.west)--(t.center)--(w.center)--(or.north);
	
	\end{tikzpicture}
	\caption{Adapted model architecture using message passing block with local and global descriptor. $\oplus$: concatenation.}
	\label{model2}
\end{figure}
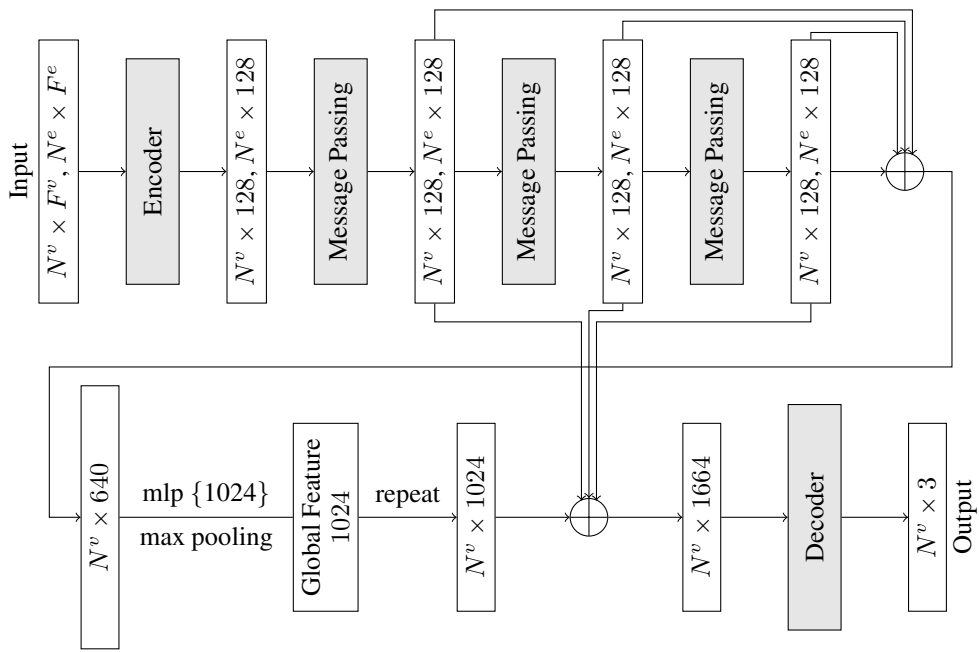
For the adaption of {\scshape{MeshGraphNets}} we are following their Encoder-Processor-Decoder architecture, but with a processor adapted by the idea of DGCNN. Instead of {\scshape{EdgeConv}} layers, message passing blocks \citep{47094} are used to update node and edge features. Figure \ref{model2} shows the visual scheme of the adapted architecture.
\textbf{Encoder} The edge feature vectors $\mathrm{\textbf{e}}_{ij}\in E$ are a concatenation of the relative displacement vector in mesh space $\mathrm{\textbf{u}}_{ij}=\mathrm{\textbf{u}}_i-\mathrm{\textbf{u}}_j$ and its norm $|\mathrm{\textbf{u}}_{ij}|$ with an edge feature vector size of $F^e$ and a total amount of edges $N^e$. The node feature vectors $\mathrm{\textbf{v}}_i\in V$ is defined by the quantities $\mathrm{\textbf{n}}_i$ and can be extended by other features with a node feature vector size of $F^v$ a total number of nodes $N^v$. In the encoder, MLPs $\epsilon^M$ and $\epsilon^V$ are used to embed the feature vectors $\mathrm{\textbf{e}}_{ij}$ and $\mathrm{\textbf{v}}_i$, respectively to latent vectors of the size 128. 
\textbf{Processor} The processor uses $L$ identical message passing blocks to update the edges $\mathrm{\textbf{e}}$ and nodes~$\mathrm{\textbf{v}}$ into embeddings $\mathrm{\textbf{e}}'$ and $\mathrm{\textbf{v}}'$ respectively by
\begin{equation}
\mathrm{\textbf{e}}'_{ij}\leftarrow f^M(\mathrm{\textbf{e}}_{ij},\mathrm{\textbf{v}}_i,\mathrm{\textbf{v}}_j)\;,\qquad \mathrm{\textbf{v}}'_i\leftarrow f^V(\mathrm{\textbf{v}}_i,\sum_j^{\phantom{x}} \mathrm{\textbf{e}}'_{ij})\;,
\end{equation}
where $f^M$ and $f^V$ are implemented as MLPs with residual connections. The MLPs are using the concatenated node and edge feature vectors as input.
\textbf{Decoder} Finally, the decoder MLP is transforming the latent node features from the last processor block into the output features. More details about this model type can be found in \citet{sanchezgonzalez2020learning} and \citet{pfaff2021learning}.
\textbf{Local and global Features} {\scshape{EdgeConv}} captures both the local neighborhood information $(\mathrm{\textbf{v}}_j-\mathrm{\textbf{v}}_i)$ and the global shape structures by $\mathrm{\textbf{v}}_i$. 
Therefore, we are adapting the node features by using a concatenation of the quantities $\mathrm{\textbf{n}}_i$ and the absolute node position $\mathrm{\textbf{u}}_i$. This node features help to capture the global shape structure, while the edge features capture the local neighborhood information. This adapted version is called {\scshape{MeshGraphNets (Abs)}}.
Further, we adopt the structure of the DGCNN architecture to increase the ability of extracting local and global properties. Instead of stacking $L$ message passing block, all the message passing blocks are concatenated to calculate the global feature vector. Then, again the local and global features are concatenated and transformed to the desired output by the decoder MLP. This adapted version is called {\scshape{MeshGraphNets (Abs+Pool)}}.
All MLPS in the encoder, message passing blocks and decoder are using three fully-connected layers of size 128. The two hidden layers come with ReLU activation, followed by a non-activated output layer. The MLPs except the decoder are followed by a LayerNorm \citep{ba2016layer} layer. Similar to our method, we are using $L=5$ message passing block.
\newpage
\subsection{Additional Results}
\begin{figure}[h]
    \begin{subfigure}{.45\textwidth}
        \centering
        \includegraphics[width=0.49\linewidth]{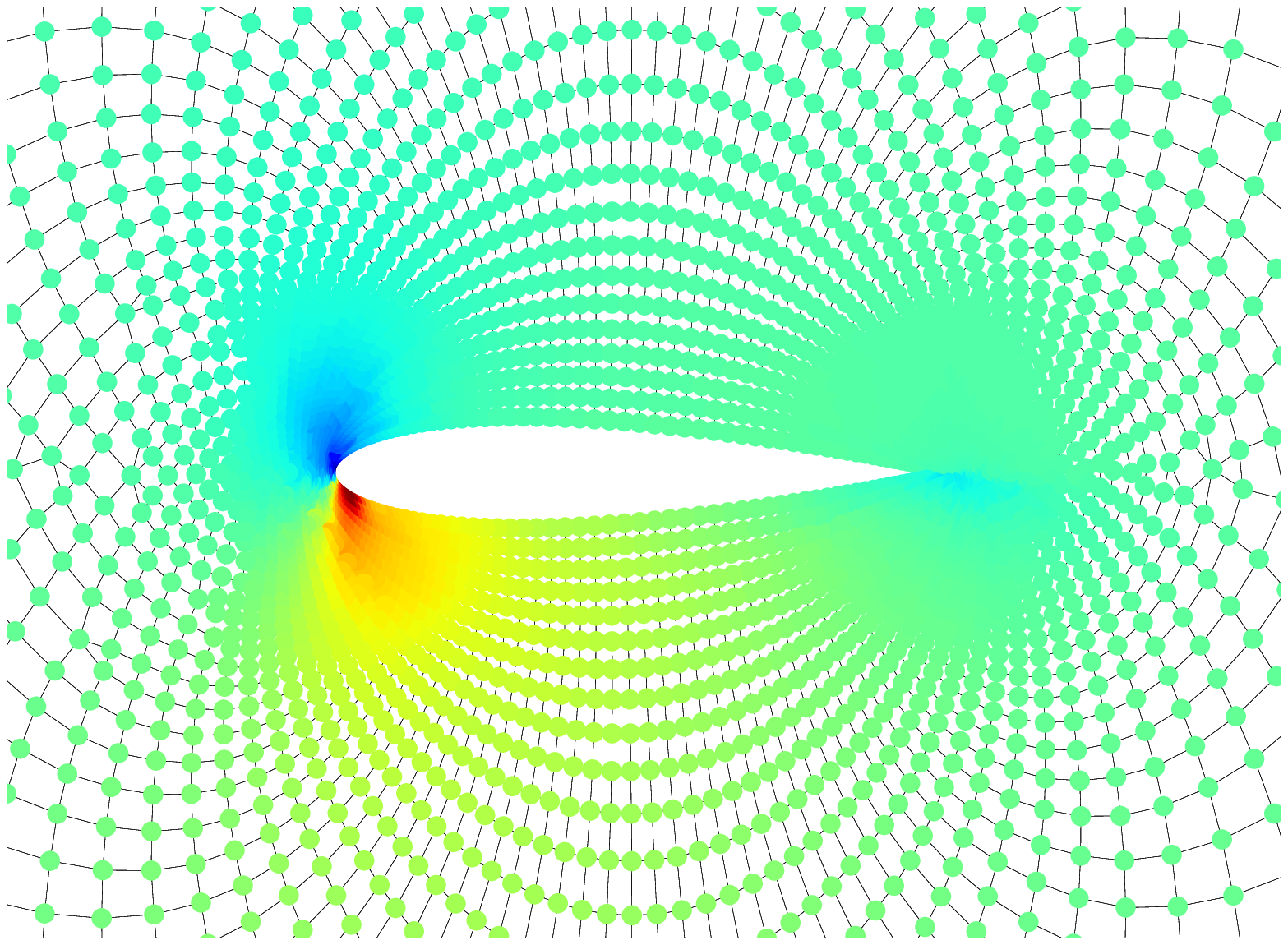}
        \includegraphics[width=0.49\linewidth]{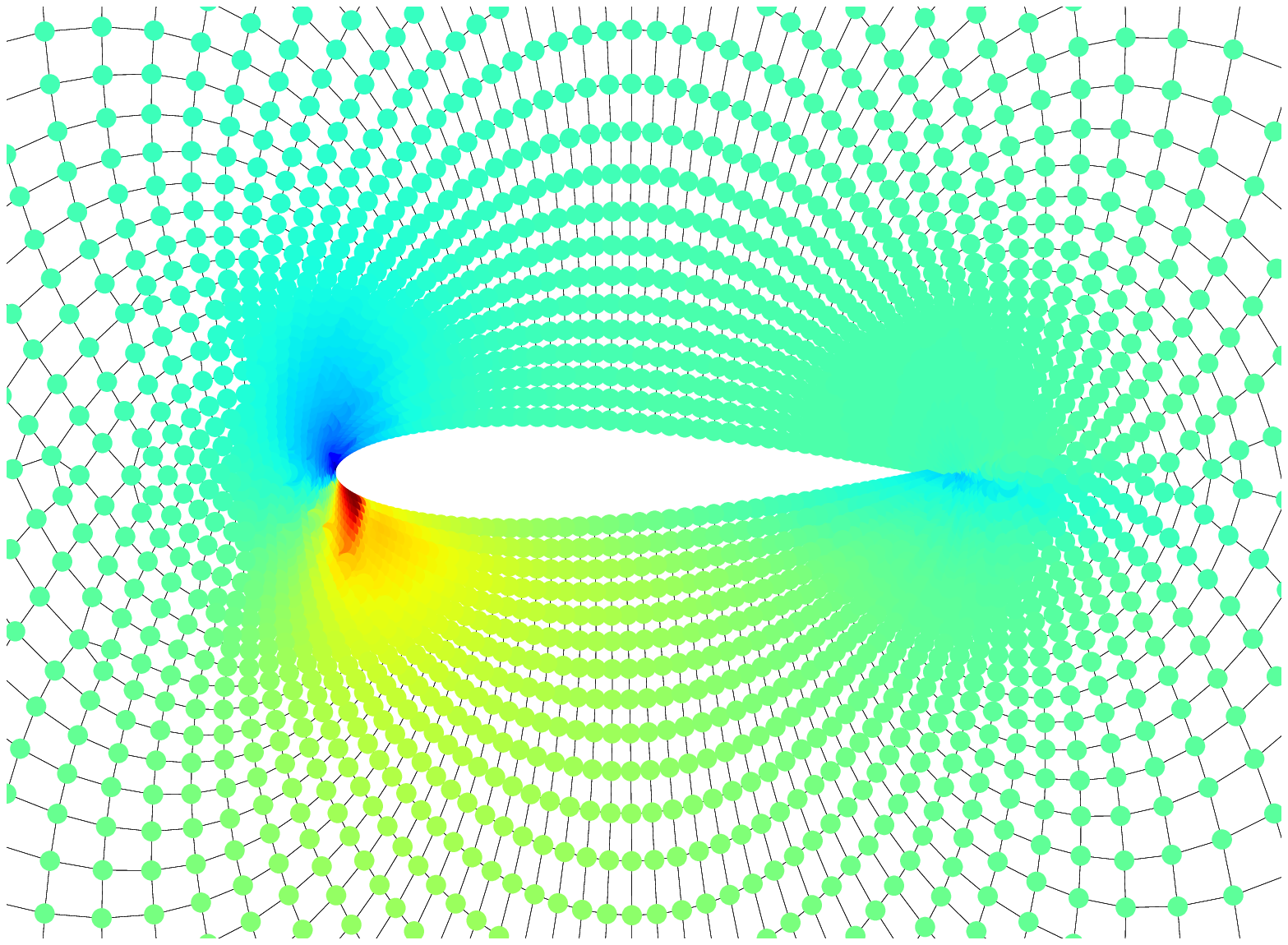}
        \caption{{\scshape{Airfoil}}}
    \end{subfigure}%
    \hfill
    \begin{subfigure}{.45\textwidth}
        \centering
        \includegraphics[width=0.49\linewidth]{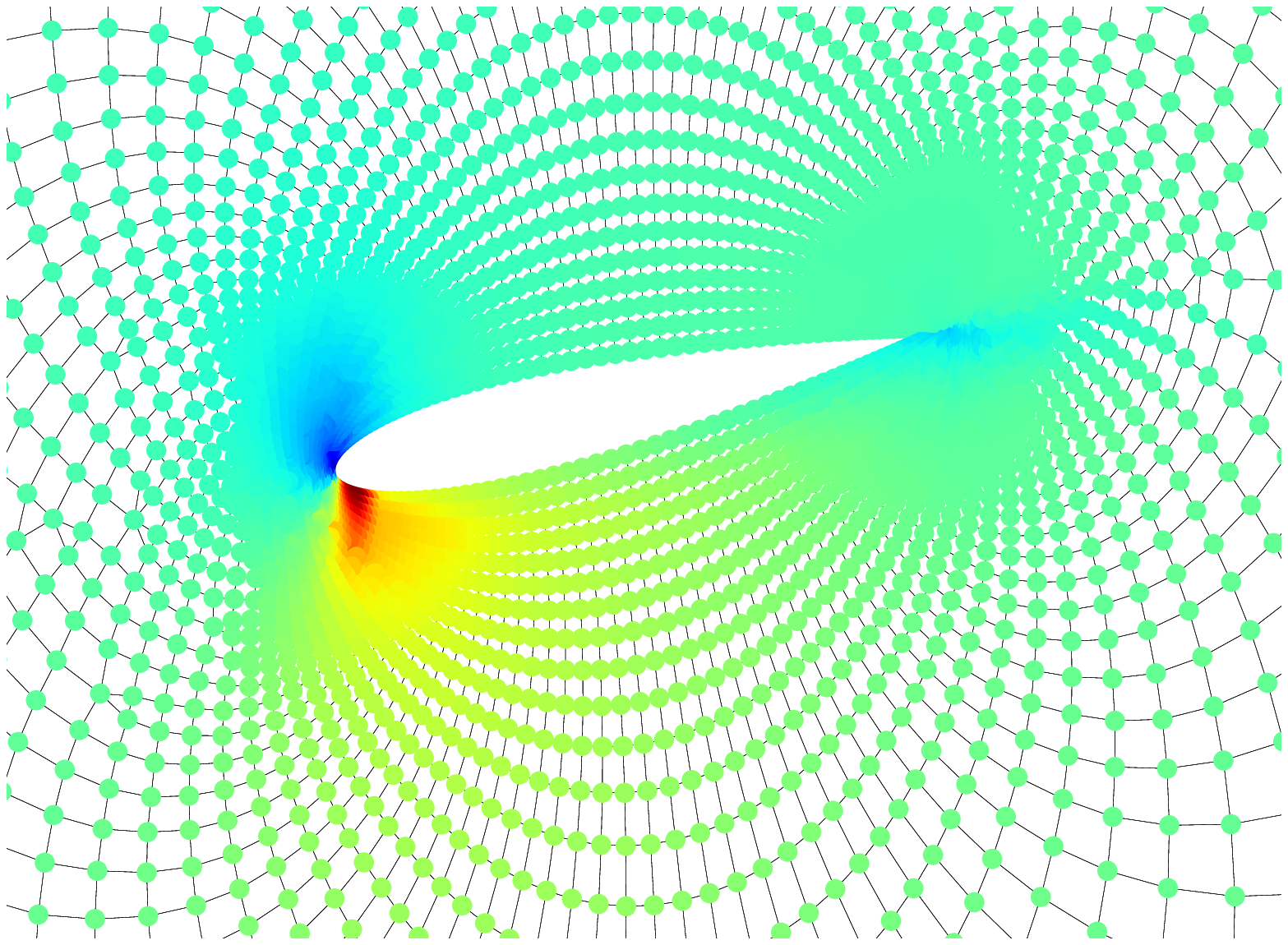}
        \includegraphics[width=0.49\linewidth]{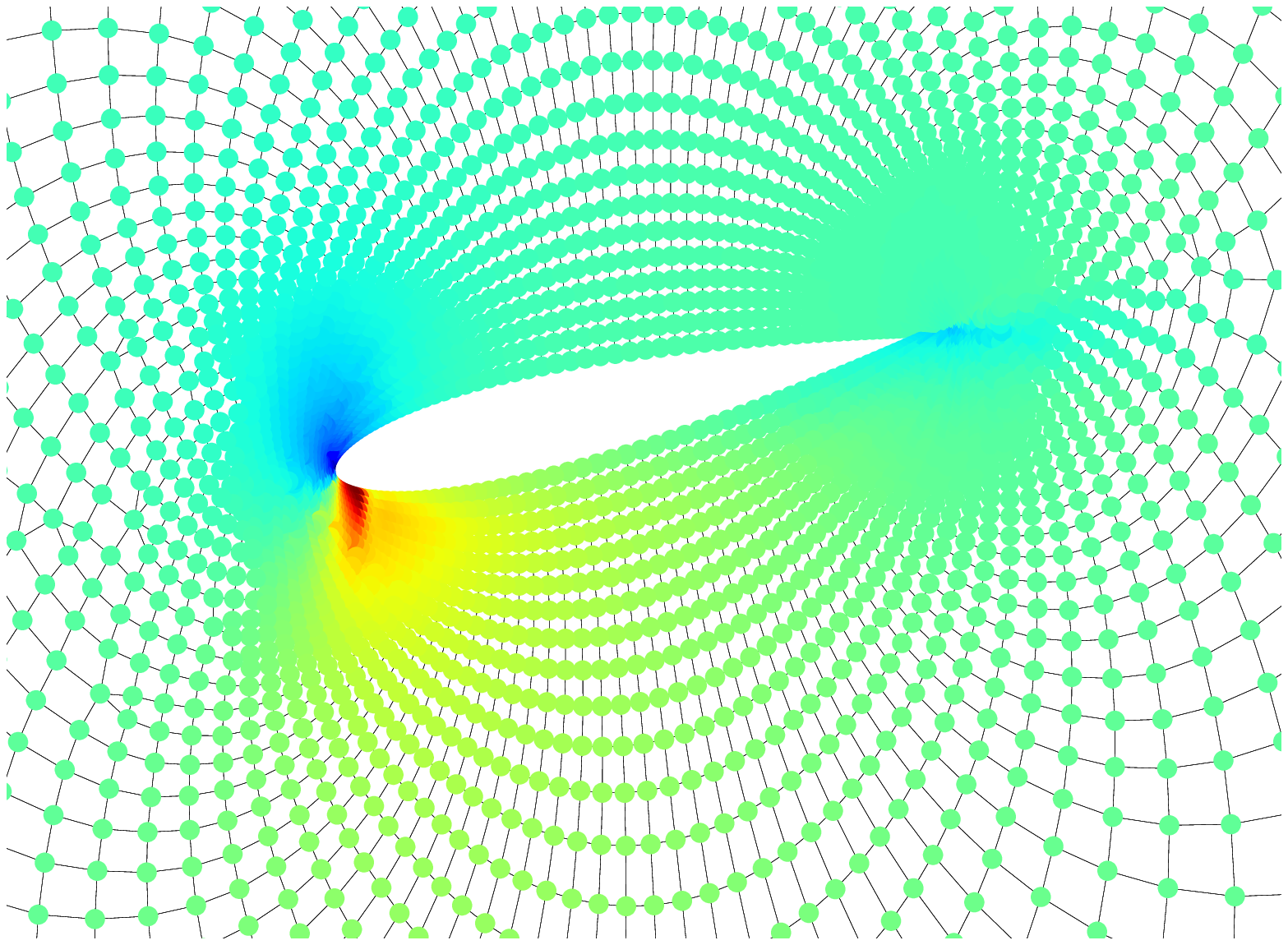}
        \caption{{\scshape{AirfoilRot}}}
    \end{subfigure}%
    \\
    \begin{subfigure}{.45\textwidth}
        \centering
        \includegraphics[width=0.49\linewidth]{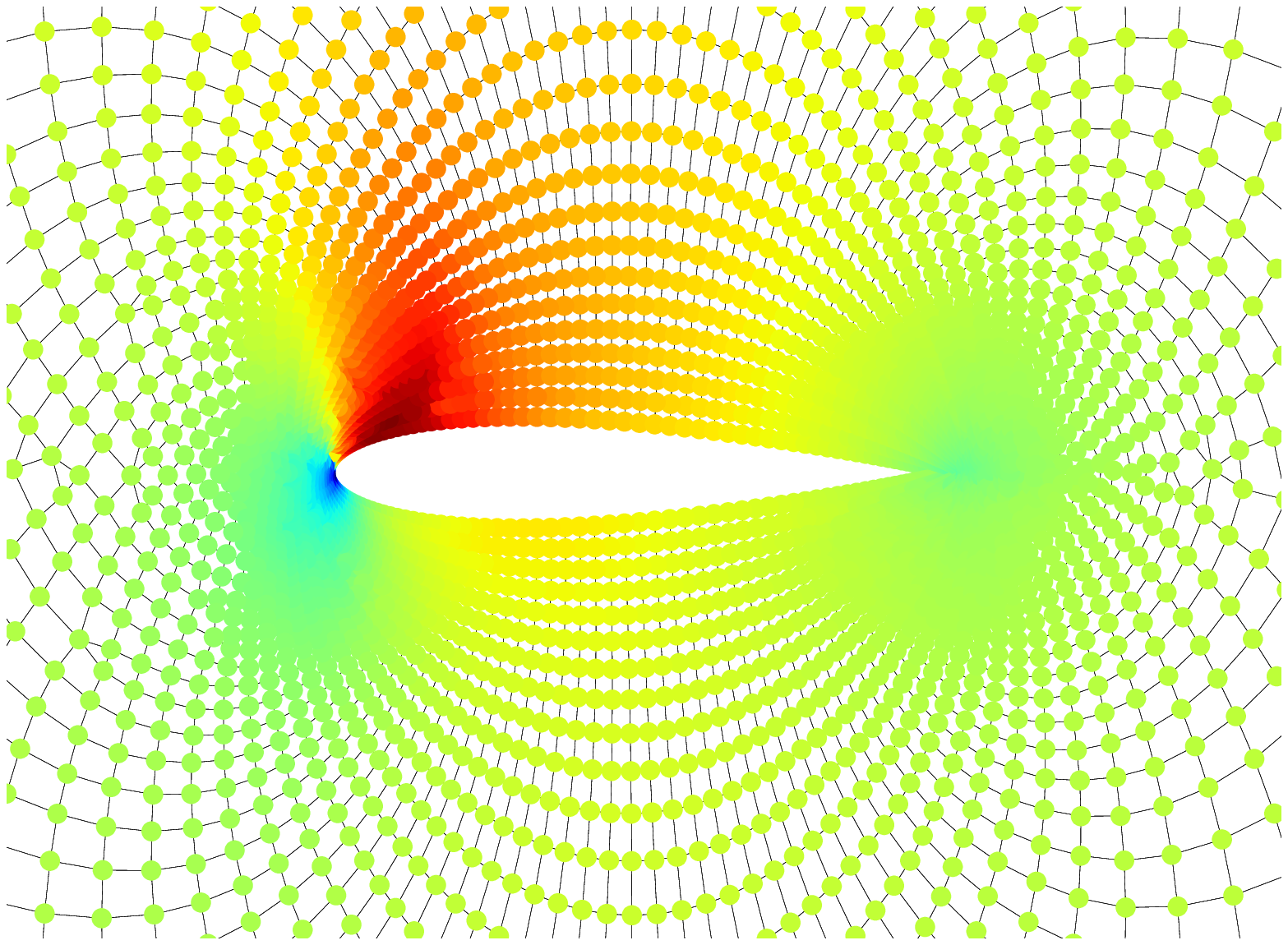}
        \includegraphics[width=0.49\linewidth]{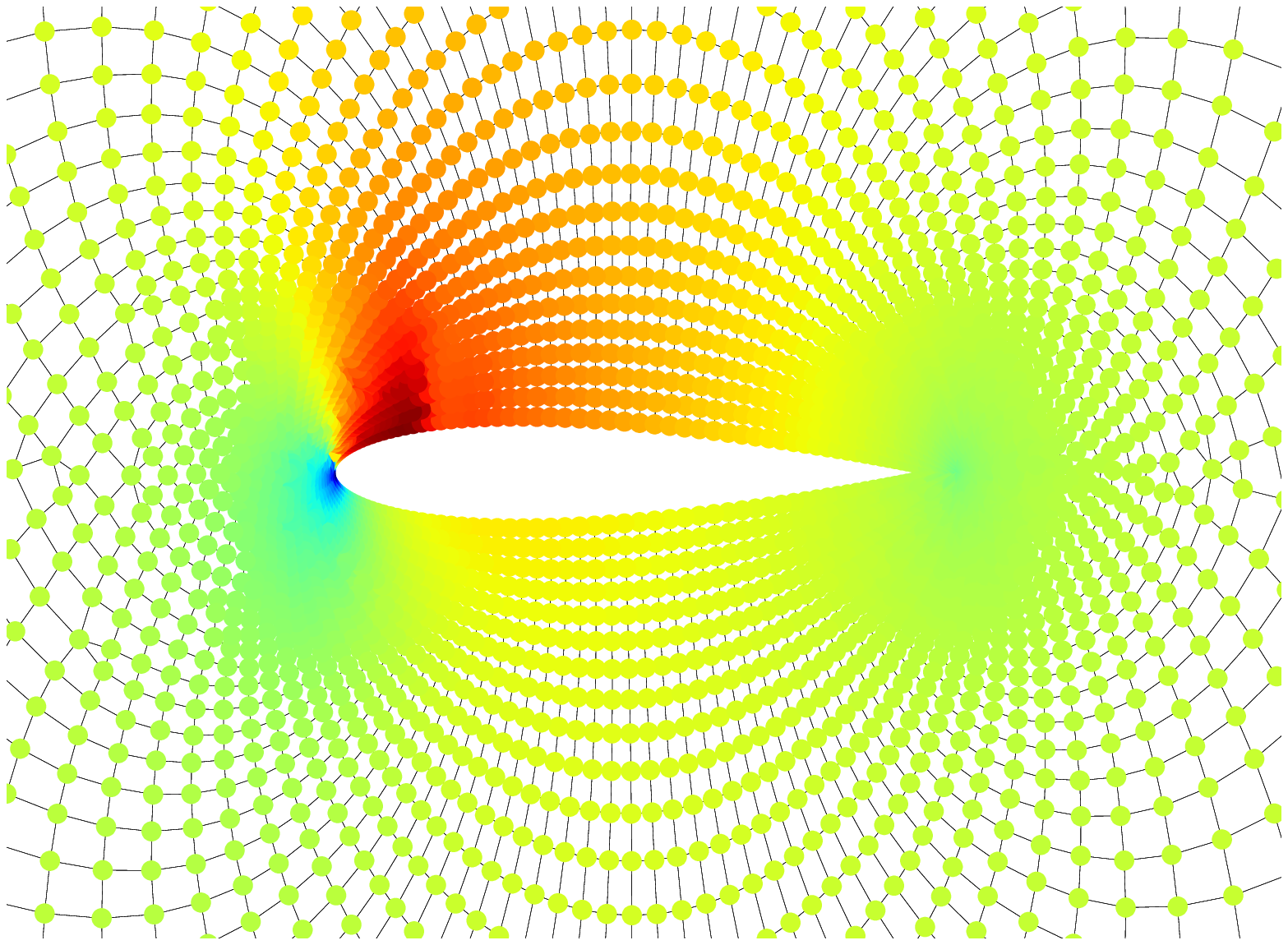}
        \caption{{\scshape{Airfoil}}}
    \end{subfigure}%
    \hfill
    \begin{subfigure}{.45\textwidth}
        \centering
        \includegraphics[width=0.49\linewidth]{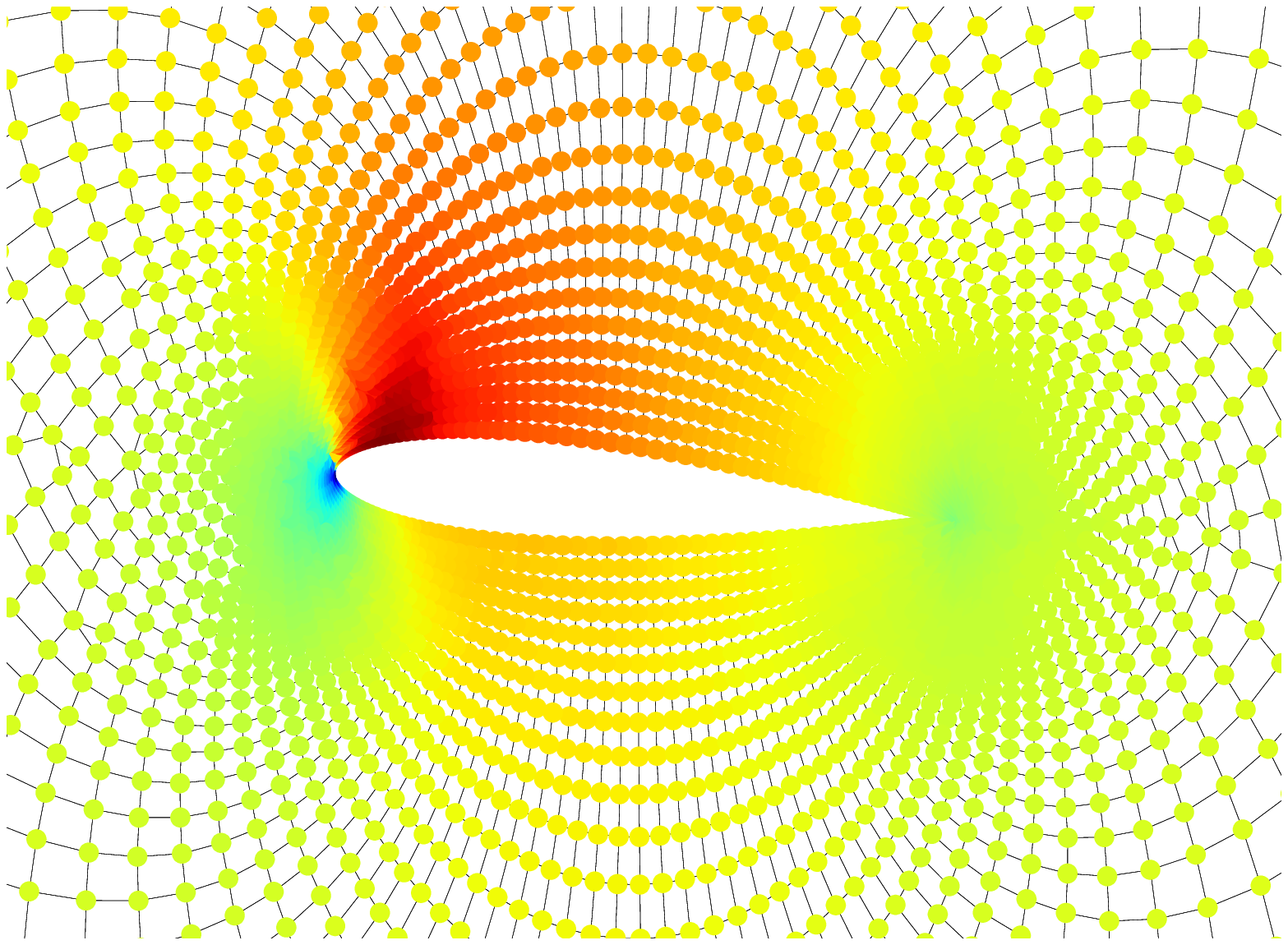}
        \includegraphics[width=0.49\linewidth]{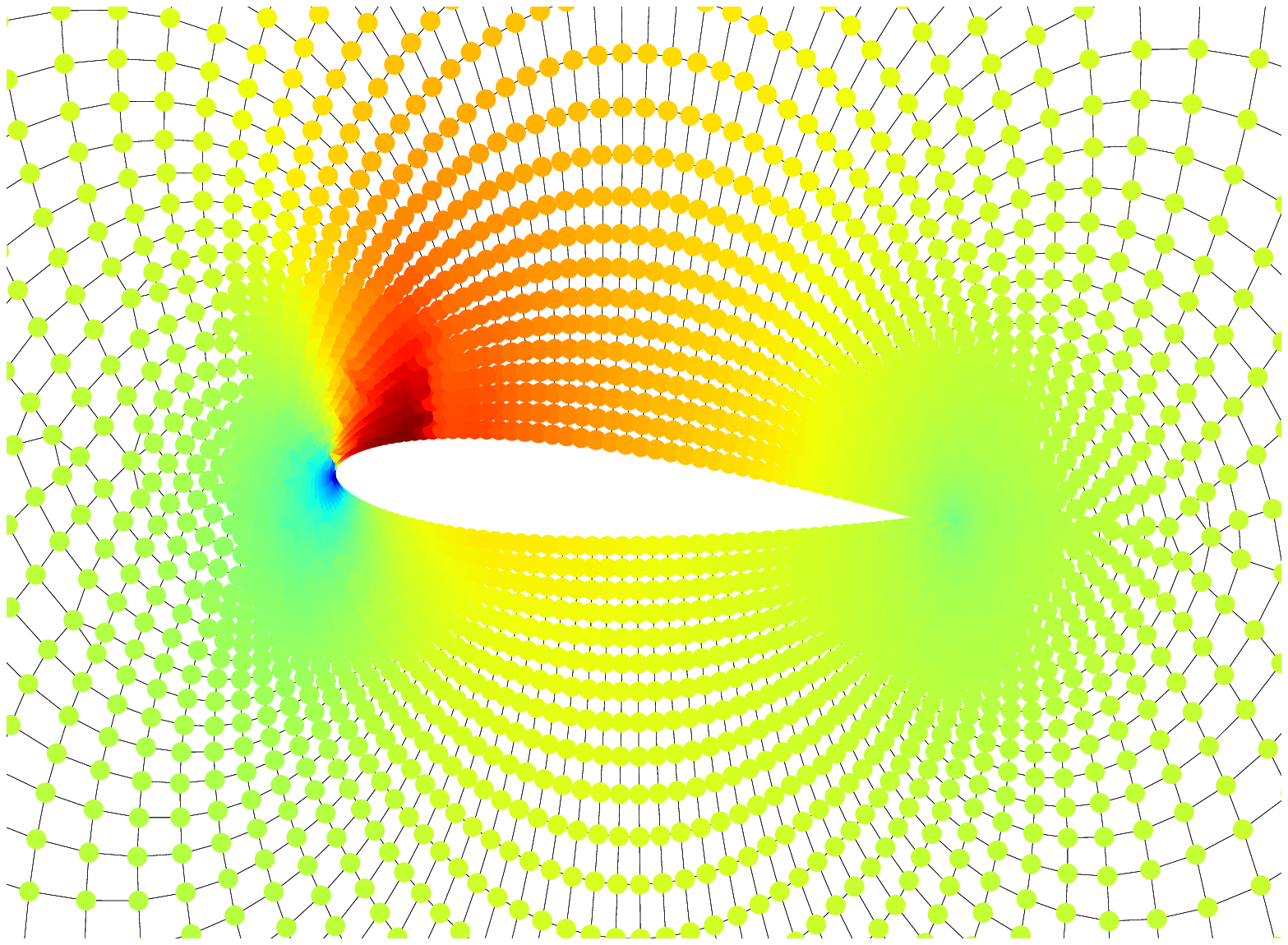}
        \caption{{\scshape{AirfoilRot}}}
    \end{subfigure}%
    \\
    \begin{subfigure}{1\textwidth}
        \centering
        \includegraphics[width=0.49\linewidth]{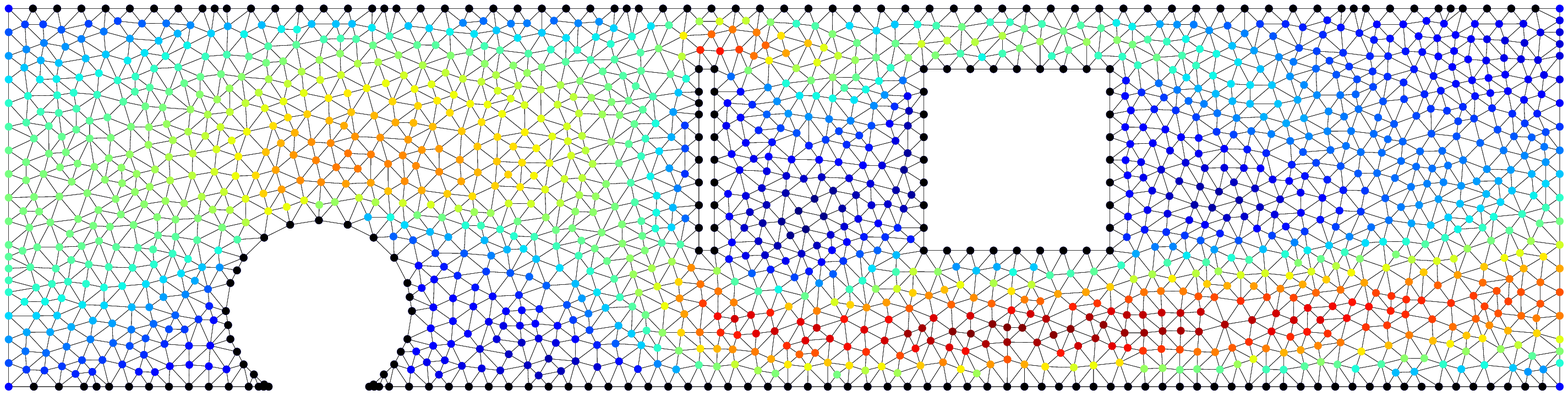}
        %
        %
        \includegraphics[width=0.49\linewidth]{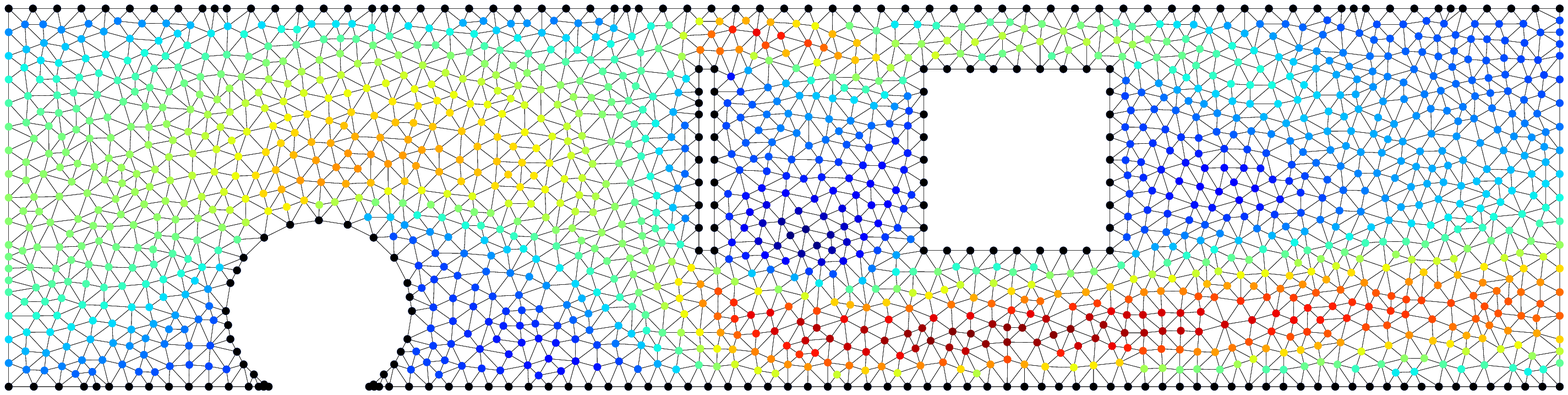}
        \caption{{\scshape{Channel}}}
    \end{subfigure}
    \\
    \begin{subfigure}{1\textwidth}
        \centering
        \includegraphics[width=0.49\linewidth]{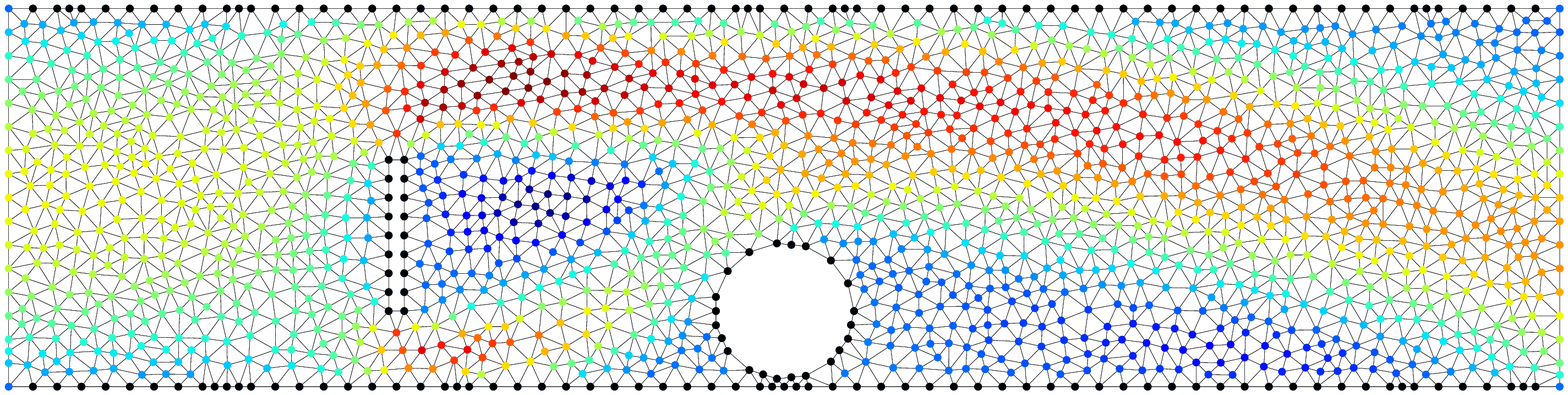}
        %
        %
        \includegraphics[width=0.49\linewidth]{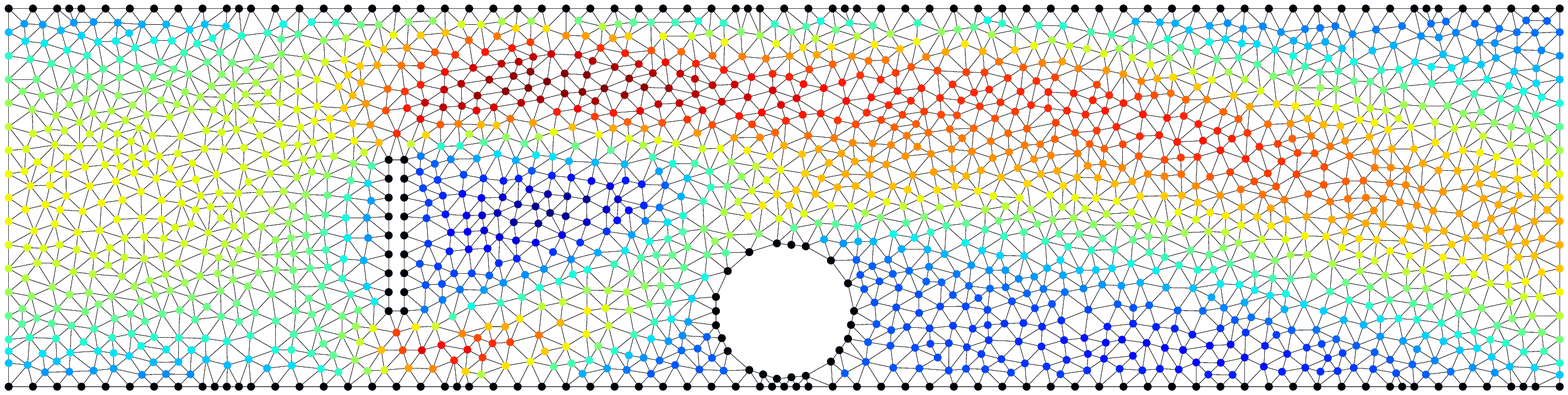}
        \caption{{\scshape{Channel}}}
    \end{subfigure}
    \\
    \begin{subfigure}{1\textwidth}
        \centering
        \includegraphics[width=0.49\linewidth]{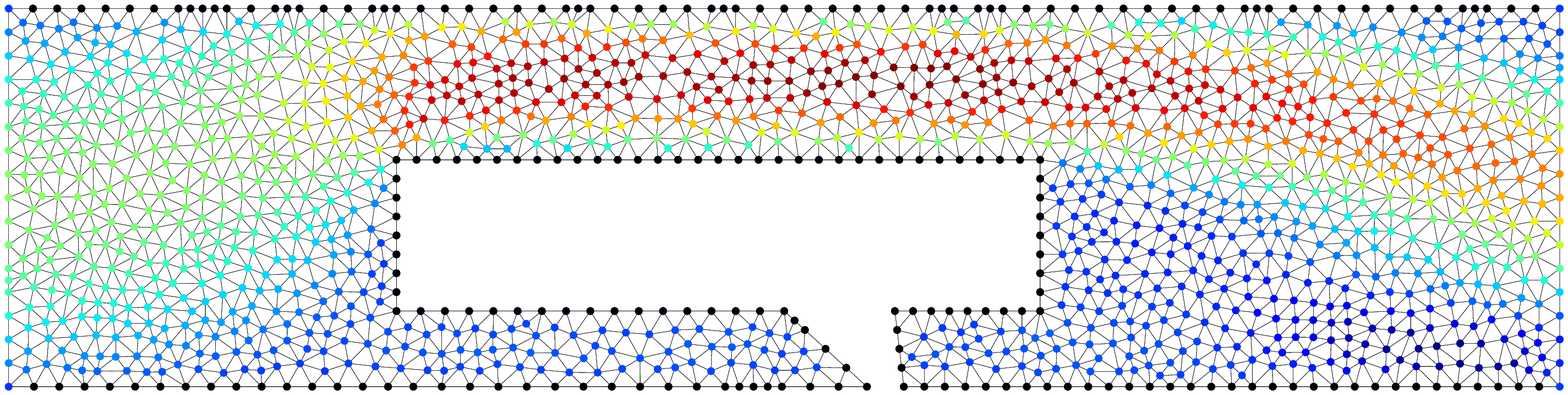}
        %
        %
        \includegraphics[width=0.49\linewidth]{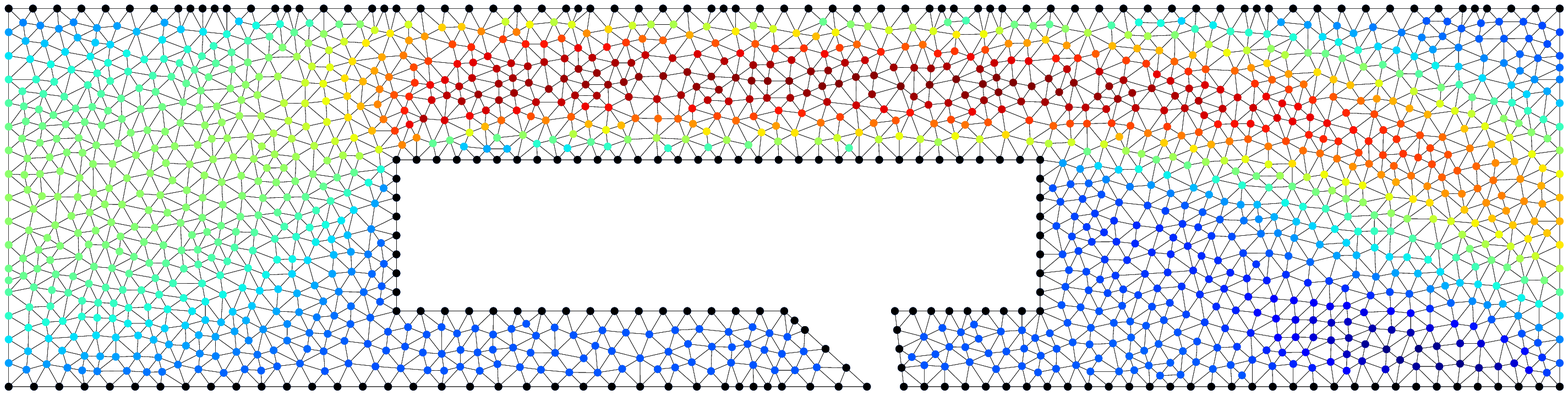}
        \caption{{\scshape{Channel}}}
    \end{subfigure}
    \caption{Additional examples of the predicted velocity field in $x$-direction for all three data sets, each with the ground truth simulation to the right.}
    \label{resultsDetails}
\end{figure}
\end{document}